\documentstyle[12pt,aps,prl,floats,eqsecnum,psfig,euscript]{revtex}
\def\lapprox{\,\raise0.4ex\hbox{$<$}\kern-0.8em\lower0.7ex\hbox{$\sim$}\,}
\def\gapprox{\,\raise0.4ex\hbox{$>$}\kern-0.8em\lower0.7ex\hbox{$\sim$}\,}
\begin{document}
\title{Spin relaxation in a two-electron  quantum dot }
\medskip
\author{{S.
Dickmann${}^{1,2}$}${}\!$\renewcommand{\thefootnote}{\fnsymbol{footnote}}
\footnotemark[1]
 and P. Hawrylak${}^3$}
\address{
${}^1$Institute for Solid State Physics of Russian Academy of
Sciences,\\
142432 Chernogolovka, Moscow District, Russia\\}
\address{
${}^2$Max-Planck-Institut f\"ur Physik Komplexer Systeme,\\
N\"othnitzer Str. 38, D-01187 Dresden, Germany\\}
\address{${}^3$Institute for Microstructural Science, National
Research Council,\\
Ottawa, Canada K1A 0R6\\
\parbox{14.2cm}
{\rm
\bigskip
\noindent
We discuss the rate of  relaxation of the total spin
 in the
two-electron droplet in the vicinity of the
 magnetic field driven singlet-triplet transition.
 The total spin relaxation is attributed to spin-orbit and electron-phonon
 interactions. The relaxation process is found to depend on the
  spin of ground and excited states. This asymmetry  is used
 to explain puzzles in recent
 high source-drain transport experiments.
\vspace{-2mm}
\begin{flushleft}
\end{flushleft}}}
\maketitle
\renewcommand{\thefootnote}{\fnsymbol{footnote}}
\footnotetext[1]{Fax: +7 096 524 9721; e-mail: dickmann@issp.ac.ru}
\vspace{90mm}

\newpage

The electronic structure of artificial atoms  in
quantum dots (QD's)\cite{dot reviews}
can be directly studied in transport measurements. Such
voltage-tunable experiments in a varying magnetic field (e.g., see Refs.
 \cite{dot reviews,ray secs,pawel prb 1999,pawel secs,sa01,fu02,ky02} and
 the works cited therein) enable to change
the number of electrons confined in a QD as well as the mutual arrangement of
electronic energy levels. The arrangment of levels depends on the total spin
of electronic configurations\cite{pawel prb 1999,pawel secs}.
The application of
the magnetic field induces transitions in the ground state between states
characterised by different total spin.  In this paper we focus on a
role played by total spin
in a simplest system, a two-electron droplet. At low magnetic fields the 
ground state
is a singlet with total spin $S=0$ while at higher magnetic fields the 
ground state is
a triplet with total spin $S=1$,
 in analogy to the
parahelium-orthohelium transition ,  already studied
theoretically in Refs.\cite{pawel secs,wa92} . Both singlet and triplet states
and the singlet-triplet (ST)
transition have been observed experimentally
 in both vertical \cite{ray secs,fu02}  and
lateral QD's \cite{sa01,ky02}.  However, while at least one of the triplet 
excited state is observed
for magnetic fields below the ST transition, the singlet excited state
is not seen past the ST transition, resulting in asymmetric (in $B$)
high source and drain transport spectra.
In the present work we present model calculations of total spin relaxation
due to the mixture of spin-orbit (SO)
 and electron-phonon interaction which helps to explain
  the unusual behaviour of the levels associated with
different total spin seen in transport experiments.

We start with
the SO interaction Hamiltonian for a
two-dimensional (2D) electron in a quantum well  written
as \cite{an82,dy86,ba92,di96,di00}:
$$
  H_{SO}=\alpha\left({\bf k}\times\hat{\mbox{\boldmath $\sigma$}}
  \right)_z-
  \beta\left(\overline{\bf k}\cdot\hat{\mbox{\boldmath $\sigma$}}
  \right)\,,         \eqno (1)
$$
where the layer plane is  determined by the principal
axes $(x,y)$
of the crystal. This expression is a combination of the  Rashba
term \cite{by84} (with the coefficient $\alpha$), and of a 2D
analogue of the Dresselhaus term (with the coefficient $\beta$)\cite{dy86,ba92}.
 We use the following notations:
${\bf k}=-i{\bf \nabla}+e{\bf A}/c$ and  $\overline{\bf k}=(k_x, -k_y)$
are 2D vectors; $\hat{\sigma}^{x,y}$ are the Pauli matrices.
The $\beta$ coefficient is determined by the
formula \cite{dy86,ba92}:
$$
  \beta=\gamma_{\mbox{\scriptsize so}}\frac{\hbar}{\sqrt{2m^*G}}\cdot
  {\mbox{Ry}}^*
  \cdot a_0^2/d^2\equiv {\cal B}\cdot{\mbox{Ry}}^*\cdot a_0^3/d^2\,,
                       \eqno(2)
$$

where $G$ is the band-gap width of the semiconductor,
$\gamma_{\mbox{\scriptsize so}}$ is the spin-orbit constant
\cite{pi84}, $\,a_0=\hbar^2\varepsilon/m^*e^2$ is the effective
Bohr radius. (For GaAs: $G=1.52\,$eV, $\gamma_{\mbox{\scriptsize so}}=0.07$,
$a_0=9.95\,$nm Ry${}^*\approx 5.74\,$meV, $m^*=0.067m_e$, and
${\cal B}=0.0043$.)
The parameter $d$ is determined  by averaging the
square of the wave-vector ${\hat z}$-component of a 2D electron in the
layer:
$
  d^{-2}=-\left\langle f\left|d^2/dz^2\right|f\right\rangle,
$
where $f(z)$ is the corresponding size-quantized function. (In the well we
consider that $f=(2/\pi)^{1/4}e^{-z^2/d^2}/\sqrt{d}$.)

The Hamiltonian of the system of two interacting electrons in the harmonic
potential $\frac{1}{2}m^*\omega_0^2(x^2+y^2)$ may be
written in Center-of-Mass (CM) and relative coordinates as
$
  {\cal H}=H_m({\bf r})+{\EuScript H}_M({\bf R}) +{\cal H}_{S}$.
Here ${\bf R}=({\bf r}_1+{\bf r}_2)/2$  and ${\bf r}= {\bf r}_1-{\bf r}_2$
are Center-of-Mass and relative (Jacobi) coordinates
  of two particles.
The first term in the Hamiltonian is
$$
  H_m=-\frac{\hbar^2}{2\mu}\left(\frac{\partial^2}{\partial r^2}+
  \frac{1}{r}\frac{\partial}{\partial r}-\frac{m^2}{r^2}\right)-
  \frac{\hbar}{2}\omega_cm+
  \frac{\mu r^2\omega_0^2}{2}\left(1+\frac{h^2}{4}\right)+
  e^2/\varepsilon r  \eqno(3)
$$
(here $\omega_c=eB/m^*c$; $\mu=m^*/2$ is the
reduced mass, $h=\omega_c/\omega_0$ is the dimensionless magnetic field).
The expression for ${\EuScript H}_M({\bf R})$ may be found from
Eq. (3) for $H_m$
after the substitution: ${\bf r}\to{\bf R},\quad \mu\to{\cal M}=2m^*,
\quad m\to M,$ and $\varepsilon\to \infty$.

The term ${\cal H}_{S}$ is the spin-dependent part of the Hamiltonian,
namely it is a combination of the Zeeman and spin-orbit coupling terms:
${\cal H}_{S}=\sum_{i=1,2}\left[|g\mu_BB|{\hat \sigma}^{z}_{i}+
H^{(i)}_{SO}\right]$. For two electrons, the SO part
$H_{SO}^{(1)}+ H_{SO}^{(2)}$ can also be
written in CM and relative  coordinates ${\bf R}$ and ${\bf r}$.
Denoting
$
  \hat{\mbox{\boldmath $\Sigma$}}=
  \hat{\mbox{\boldmath $\sigma$}}_{1}+\hat{\mbox{\boldmath $\sigma$}}_{2},
  \quad
  \hat{{\mbox{\boldmath $\sigma$}}}=
  \hat{\mbox{\boldmath $\sigma$}}_{1}-
                   \hat{\mbox{\boldmath $\sigma$}}_{2}\,.
$
we obtain
$$
  {\cal H}_{S}=\left(H_{RSO}+ H_{rSO}\right) + g\mu_BB\hat{\Sigma}_z,
  \eqno (4)
$$
where
$$
   H_{RSO}=-D_+\left(\alpha\hat{\Sigma}_-+i\beta\hat{\Sigma}_+\right)-
   D_-\left(\alpha\hat{\Sigma}_+-i\beta\hat{\Sigma}_-\right)\,,
$$
and
$$
   H_{rSO}=-{\partial}_+\left(\alpha\hat{{\sigma}}_-+
   i\beta\hat{{\sigma}}_+\right)-
   {\partial}_-\left(\alpha\hat{{\sigma}}_+-
   i\beta\hat{{\sigma}}_-\right)\,.  \eqno(5)
$$
The new operators are
$$
  D_{\pm}=\mp\frac{1}{2}\left(\frac{\partial}{\partial X}\pm
  i\frac{\partial}{\partial Y}\right)+b(X\pm iY),\qquad
  \partial_{\pm}=\mp\left(\frac{\partial}{\partial x}\pm
  i\frac{\partial}{\partial y}\right)+\frac{b}{2}(x\pm iy)
$$
[$(X,Y)$ and $(x,y)$ are the  components of 2D vectors
${\bf R}$ and ${\bf r}$] , $b=m^*\omega_c/2\hbar$ and
$$
  \hat{\Sigma}_{\pm}=\frac{1}{2}\left(\hat{\Sigma}_x\pm i\hat{\Sigma}_y\right),
  \qquad
  \hat{{\sigma}}_{\pm}=\frac{1}{2}\left(\hat{{\sigma}}_x\pm
  i\hat{{\sigma}}_y\right)\,.
$$

The wavefunction of  the two-electron
system may be written  in the form $\Psi_M(R)e^{iM\Phi}\cdot\psi_m(r)e^{im\phi}
\chi_{\sigma_1,\sigma_2}$, where $\sigma_1$ and  $\sigma_2$ are the spin
variables of the electrons. We expand the wavefunction in the basis
 set of the singlet and triplet states
$$
  |\mbox{s}\rangle=\Psi_0(R)\psi_0(r)|0,0\rangle\quad\mbox{and}\quad
  |\mbox{t},S_z\rangle=\Psi_0(R)\psi_1(r)e^{i\phi}|1,S_z\rangle\quad
  (S_z=0,\pm 1)\,.           \eqno (6)
$$
Here $\Psi_0$ is the ground state function (i.e. it obeys
the equation ${\EuScript H}_0\Psi_0=\hbar\omega_0\sqrt{1+h^2/4}\Psi_0$),
while the functions
$\psi_{0,1}$ have to be found from the equations
$$
  H_m\psi_m=E_m\psi_m\,.  \eqno (7)
$$

The analytical solution of Eqs. (7) could be found if
$l=\sqrt{\hbar/m^*\omega_0}(h^2+4)^{-{1}/{4}}\ll a_0$ or $l\gg a_0$.
The $l\ll a_0$ case has been studied perturbatively in Ref.
\cite{wa92}. Here we consider the opposite limit $l\gg a_0$
(this seems to be more relevant to a typical experimental
situation).
Then in the
leading approximation the solutions of Eqs. (7) are the states of
one-dimensional oscillator with mass $\mu$ and frequency
$\omega_0\sqrt{3+3h^2/4}$ localized in the vicinity of
$r_0=\left[{2l_0^4}/{a_0(1+h^2/4)}\right]^{1/3}\gg l_0\,$ (here we have
designated
$l_0=\sqrt{\hbar/m^*\omega_0}$). In this approximation the energy
measured from the
ground state is
$$
  \delta_{m,S_z}=E_m-E_0=\frac{\hbar^2m^2}{r_0^2m^*}-
  \frac{\hbar}{2}\omega_c\left(m+g^*S_z\right)\equiv
  \frac{\hbar\omega_0}{4}D_{m,S_z}(h)                 \eqno (8)
$$
($g^*=gm^*/m_e\approx 0.029$).
The equation determining the ST crossing takes thereby the form
$\delta_{1,S_z}=0$ (if $a_0/l_0=1/3$, then for the ST crossing point
we obtain $h=0.64$ at $S_z=0$, whereas
for the same $a_0/l_0$ and $S_z$ the exact numerical calculation \cite{wa92}
gives $h=0.69$).

We now turn to the effect of SO interaction on the mixing of
singlet and triplet states.
Operators $\Sigma_{\pm}$ and $\Sigma_z$ commute with
${\mbox{\boldmath $\Sigma$}}^2$, therefore the first and third terms in
Eq. (4) are not responsible for mixing of the singlet and triplet states.
On the contrary, the second term in Eq. (4) results in this mixing.
Indeed, let
$|S,S_z\rangle$ be the normalized spin states of two electrons. When
operating on the state $|0,0\rangle$, the term $H_{rSO}$ yields
the following non-zero matrix elements:
$
  \langle 1,1|H_{rSO}|0,0\rangle$ and  $\langle 1,-1|H_{rSO}|0,0\rangle$.
(One can check that $\hat{{\sigma}}_{\pm}|0,0\rangle=\mp2^{-1/2}|1,
\pm 1\rangle$.)

Hence we see immediately the $|0,0\rangle$ singlet state is mixed with
the $ | 1,\pm 1 \rangle$ triplet states but not with the $ | 1,0 \rangle$
state. This state is therefore long-lived.

For the states which are coupled, the expansion in terms of the small
parameter $a_0/l_0$ leads to the following result for the mixing matrix element between
$|s\rangle$ and $|t,1\rangle$ states:
$$
  M_{SO}=\langle s|H_{rSO}|t,1\rangle \approx
  -\frac{i\beta}{2^{3/2}}\left(\frac{1}{r_0}+{br_0}\right)
   \equiv -\frac{i\beta}{2^{3/2}l_0}{\cal L}(h)\,.  \eqno(9)
$$
If we take into account the Rashba term, we would find
that another non-zero matrix element is $\langle s|H_{rSO}|t,-1\rangle$,
however {\it the state $|t,0\rangle$ is never mixed with the singlet}. Further
we neglect the Rashba coupling (usually there is $\alpha<\beta$ in GaAs
hetero-structures, besides $\alpha$ vanishes in the case of
an ideally symmetric quantum well).

With help of Eq. (9) we find the hybridized states
$|\mbox{S}\rangle=C^-_0|s\rangle+C^-_1|t,1\rangle$, and
$|\mbox{T}\rangle=C^+_0|s\rangle+C^+_1|t,1\rangle$, where
$$
  C^{\pm}_0=\left\{\frac{M_{SO}}{2\left|M_{SO}\right|}\left[1\mp
  \frac{\delta_{1,1}}{\delta\!E }\right]\right\}^{1/2},\,\qquad
  C^{\pm}_1=\pm\left\{\frac{M_{SO}}{2\left|M_{SO}\right|}\left[1\pm
  \frac{\delta_{1,1}}{\delta\!E }\right]\right\}^{1/2}   \eqno(10)
$$
$\left(\delta\!E =\sqrt{\delta_{1,1}^2+4\left|M_{SO}\right|^2}\right)$.
The corresponding energies are
$E_{\mbox{\scriptsize T}/\mbox{\scriptsize S}}=
E_0+\left(\delta_{1,1}\pm\delta\!E \right)/2$.

The next step in our study is the calculation of the $|\mbox{T}\rangle\!
\to\!|\mbox{S}\rangle\;\;$ (or $|\mbox{S}\rangle\!\to\!|\mbox{T}\rangle$)
relaxation rate for the case, where $\delta_{1,1}>0\;\;$
(or $\delta_{1,1}<0$). Evidently
the main relaxation channel is determined by emission of the
acoustic phonon with energy
$\hbar c_s{k}=\delta\!E $; where
${\bf k}=({\bf q},k_z)$ is the phonon wave
vector, and $c_s$ is the mean sound velocity (we use the so-called
isotropic model, i.e., $c_s$ does not depend on the polarization and
on the ${\bf k}$ direction; we consider that $c_s=3.37\cdot 10^5\,$cm/s).
The probability of this event is
determined by
$$
  \frac{1}{\tau}=\sum_{{\bf q},k_z}\frac{2\pi
  \left|{\EuScript M}({\bf q},k_z)\right|^2}{\hbar}
  \delta(\hbar c_s{k}-\delta\!E )\,,              \eqno (11)
$$
where ${\EuScript M}({\bf q},k_z)$ is the appropriate
matrix element
$$
  {\EuScript M}({\bf q},k_z)=\langle\mbox{T}|U_{\scriptsize{\mbox{e-ph}}}|
  \mbox{S}\rangle
  = {C^+_0}^*C_0^-\left(\langle s|U_{\scriptsize{\mbox{e-ph}}}|s\rangle-
  \langle 1,t|U_{\scriptsize{\mbox{e-ph}}}|t,1\rangle\right)\,. \eqno(12)
$$
Here the phonon field is  averaged over the angle
$\phi=r\!\wedge\! q$:
$$
  U_{\scriptsize{\mbox{e-ph}}}({\bf R},{r})=
  \left(\frac{\hbar}{V}\right)^{1/2}\sum_{s}
  {\tilde U}_s({\bf q},k_z)e^{i{\bf qR}}
  \overline{\left(e^{i{\bf qr}}+e^{-i{\bf qr}}\right)}=
  2\left(\frac{\hbar}{V}\right)^{1/2}\sum_s
  {\tilde U}_s({\bf q},k_z)e^{i{\bf qR}}J_0(qr)   \eqno(13)
$$
[$s$ is the polarization, $V$ is the sample volume, and
${\tilde U}_s({\bf q},k_z)$ is the renormalized vertex which includes the
deformation and piezoelectric fields created by the phonon]. The integration with
respect to $z$ has already been performed, and reduces to the
renormalization
${\tilde U}_s=U_s({\bf q},k_z)F(k_z)$, where the formfactor is
$F(k_z)=\langle f|e^{ik_zz}|f\rangle$.
By using Eqs. (6), (12) and the expansion
$J_0(qr)\approx J(qr_0)-q(r-r_0)J_1(qr_0)$, we obtain the matrix element
(12); and after the substitution into Eq. (11) we find that
the relaxation rate is proportional to the
$|\sum_s U_s|^2$. The latter is represented as \cite{di96}
$$
  \left|\sum_s U_s({\bf q},k_z)\right|^2=\frac{\pi\hbar c_s k}
  {p_0^3\tau_A}\,,\quad\mbox{where}\quad\frac{1}
  {\tau_A({\bf k})}=
  \frac{1}{\tau_D} +
  \frac{5p_0^2}{k^6\tau_P}(q^2k_z^2+q_x^2q_y^2)\,.   \eqno (14)
$$
The summation involves  averaging over the directions of
the polarization unit vectors for both components of the electron-phonon
interaction.
The nominal times for the deformation and piezoelectric interactions
in GaAs are $\tau_D\approx 0.8\,$ps, and $\tau_P\approx 35\,$ps
\cite{di96,gale87}. The nominal momentum is
$p_0=2.52\cdot 10^6/$cm \cite{di96,gale87}.
We also refer to Refs. \cite{di96,gale87} for details concerning
the meaning and the expressions of these values in terms of the GaAs
material parameters.

Finally, with the help of Eqs. (2),(9),(10),(12)-(14)
and (11) we calculate the relaxation time.
$$
  1/\tau(h)={\cal W}\cdot
  {\cal E}{\cal L}^2\cdot \exp{\left(-{\cal A}{\cal E}^2\right)}\cdot
{\cal J}\,.
$$
where
$$
  {\cal W}=
  \frac{\pi(\mbox{Ry})^3a_0^{10}{\cal B}^2}{216(\hbar c_s)^3p_0
  \tau_Pl_0^4d^4}\,,\qquad
  {\cal E}(h)=\frac{4\delta}{\hbar\omega_0}=\sqrt{{\cal D}_{1,1}^2\!(h)+
  2\left[{\cal B}
  a_0l_0{\cal L}\!(h)/d^2\right]^2}\,
$$
  [see the definitions for ${\cal L}$ and ${\cal D}_{1,1}$ in Eqs. (8-9)],
$$
  \!{\cal A}=
  \left(\frac{da_0^2\mbox{Ry}}{4l_0^2\hbar c_s}\right)^2\,,
  {}\quad
  {\cal J}(h)=\int_0^1\frac{d\xi\,\xi}{\sqrt{1-\xi}}
  \left\{J_1[{\cal R}(\xi)]\right\}^2\left[5\xi-\frac{35}{8}\xi^2+
  {\cal S}{\cal E}^2\!(h)\right]e^{-\xi{\cal P}(h)},
$$
$${\cal R}(\xi)\!=\!{a_0\mbox{Ry}}/{\hbar c_s}
\left({a_0}/{l_0}\right)^{2/3}\sqrt{\xi}{\cal E}(h)(4+h^2)^{-1/3}\,,\quad
{\cal S}\!=\!\left({\tau_P}/{\tau_D}\right)\left({a_0^2\mbox{Ry}}/
{2l_0^2\hbar c_sp_0}\right)^2\,,
$$
and
$$
  {\cal P}(h)=\left[{a_0^2}/{l_0^2\sqrt{4+h^2}}
  -\frac{1}{2}
  \left({da_0}/{l_0^2}\right)^2\right]{\left[{\cal
  E}(h)a_0\mbox{Ry}\right]^2}/{8(\hbar c_s)^2}\,.
$$
As an illustration
Fig. 1 shows the relaxation rate as a function of the magnetic field
on the logarithmic scale (the main picture) and in the usual scale
(the inset). The relaxation time is seen to have
a sharp maximum in the vicinity of the ST crossing but constitutes a
comparatively small value (of the order of $0.1\,$mcs) in the regions
where the singlet and triplet lines are resolved. The non-monotonic
behaviour of $\tau$ on the right of the ST crossing originates from the
correlations between the wave-function characteristic distance $r_0$
and the wavelength $\hbar c_s/\delta E$ of the emitted phonon.

\begin{figure}

\centerline{\psfig{figure=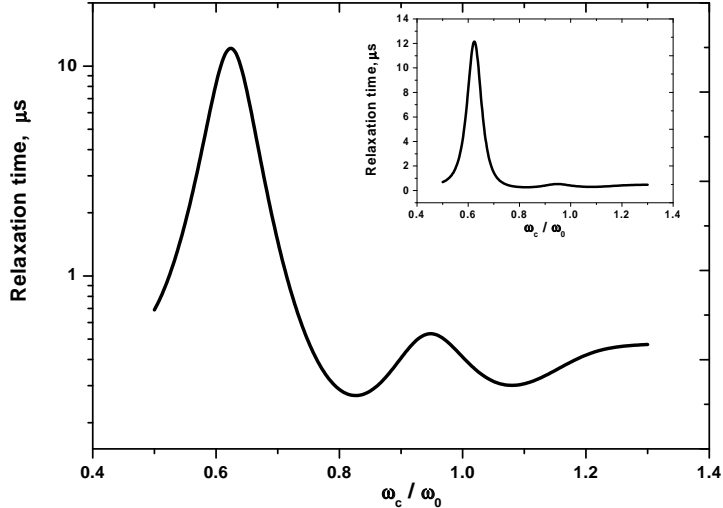,width=120mm,angle=0}}

\caption{The ST relaxation time $\tau$ calculated for $a_0/l_0=1/3$ and
$d=5\,$nm. The maximum corresponds to ST transition at  
$\omega_c/\omega_0=0.64$. See the text for details.}

$\;$
\label{Firstfig}
\end{figure}

We now turn to the discussion of the manner in which the ST relaxation
could influence the transport spectroscopy through the QD states.
By studying the kinetic processes of filling and emptying the dot
in the presence of a large ``source-drain" voltage,
we estimate the effective electron
life-time inside the dot  at the
``working level" i.e. at the level which participates in the
transport through the dot.
This effective life-time $\tau_{\mbox{\scriptsize dot}}$ under
experimental conditions of Refs. \cite{fu02,ky02} we estimate
 to be of the order of $1\,$mcs, and this value should
evidently be compared with the
ST relaxation time calculated above. If the working level is exactly
the upper level of the two-electron droplet, then the
relaxation could influence the current. Namely, if
$\tau<{\tau_{\mbox{\scriptsize dot}}}$, then the working level could be
emptied due to the ST relaxation occurring within the dot. In this case
the current through the upper two-electron state becomes negligible.
The relaxation process is asymmetric across the ST transition.
Before the transition the "working level" involves $|t,\pm 1\rangle$
and $|t,0\rangle$ [see Eq. (6)] triplet states. (The Zeeman splitting 
is not resolved.)
The $|t,0\rangle$ state is long lived and hence observed in
experiment while the $|t,1\rangle$ state relaxes efficiently to
the  $S=0$ singlet state $|s\rangle$. Past the ST transition, the 
$|t,1\rangle$ state is the ground state but the
excited state is the singlet. The singlet state relaxes efficiently
to the $|t,1\rangle$ ground state. Hence this asymmetry in the relaxation
processes associated with the singlet-triplet transition could be 
responsible for
the anomalies observed in transport experiments \cite{fu02,ky02}.

In closing it is worthy to
mention other relaxation channels which are not taken into account
in our calculation but which in the framework of  the considered
mechanism
could additionally reduce the ST relaxation time.
These are provided by special phonon modes (e.g., by surface and confinement
phonons excited in  the
hetero-junction) and certainly by the SO Rashba coupling if the latter is
significant.

The authors ackowledge A.Sachrajda, D.G. Austing and J. Kyriakidis for
discussions. S.D. acknowledges the support of the Russian
Foundation for Basic Research and thanks  the Institute
for Microstructural Science (Ottawa), where this project was initiated,
for hospitality.

\end{document}